\documentclass[dvipdfmx]{pasj01}
\draft
\usepackage{color}
\usepackage[normalem]{ulem}

\usepackage{multirow}
\usepackage[]{graphicx}

\begin{document} 

\title{The formation of the open cluster NGC~602 in the Small Magellanic Cloud triggered by colliding H{\sc i} flows}

\author{Yasuo \textsc{FUKUI}\altaffilmark{1,2}}
\author{Takahiro \textsc{OHNO}\altaffilmark{1}}
\author{Kisetsu \textsc{TSUGE}\altaffilmark{1}}
\author{Hidetoshi \textsc{SANO}\altaffilmark{3}}
\author{Kengo \textsc{TACHIHARA}\altaffilmark{1}}
\email{fukui@a.phys.nagoya-u.ac.jp}
\altaffiltext{1}{Department of Physics, Nagoya University, Furo-cho, Chikusa-ku, Nagoya 464-8601, Japan}
\altaffiltext{2}{Institute for Advanced Research, Nagoya University, Furo-cho, Chikusa-ku, Nagoya 464-8601, Japan}
\altaffiltext{3}{National Astronomical Observatory of Japan, Mitaka, Tokyo 181-8588, Japan}


\maketitle

\begin{abstract}
NGC~602 is an outstanding young open cluster in the Small Magellanic Cloud. We have analyzed the new H{\sc i} data taken with the Galactic Australian Square Kilometre Array Pathfinder survey project at an angular resolution of 30\arcsec. The results show that there are three velocity components in the NGC~602 region. We found that two of them having $\sim$20 km s$^{-1}$ velocity separation show complementary spatial distribution with a displacement of 147 pc. We present a scenario that the two clouds collided with each other and triggered the formation of NGC~602 and eleven O stars. The average time scale of the collision is estimated to be $\sim$8 Myr, while the collision may have continued over a few Myr. The red shifted H{\sc i} cloud extending $\sim$500 pc flows possibly to the Magellanic Bridge, which was driven by the close encounter with the Large Magellanic Cloud 200 Myr ago \citep{1990PASJ...42..505F,2007MNRAS.381L..11M}. Along with the RMC136 and LHA 120-N 44 regions the present results lend support for that the galaxy interaction played a role in forming high-mass stars and clusters.
\end{abstract}

\section{Introduction} \label{sec:intro}
Dynamical interactions between galaxies are a common process in the Universe and exchange of the interstellar medium (ISM) between galaxies is often happening. It is also likely that the gas motion driven by the interaction produces shocks to trigger star formation, and drives the galaxy evolution. There are a number of interacting galaxies and in the most typical case, the Antennae Galaxies, many young massive clusters are probably formed under triggering by the interactions (\cite{2000ApJ...542..120W,2014ApJ...795..156W,2019arXiv190905240T}, \yearcite{2020arXiv200504075T}).

The Magellanic Clouds are a tidally interacting system with close encounters in the past. \citet{1990PASJ...42..505F} proposed that the most recent encounter 200 Myrs ago produced H{\sc i} flows driven by the tidal force between the Large Magellanic Cloud (LMC) and the Small Magellanic Cloud (SMC). \citet{2017PASJ...69L...5F} analyzed the H{\sc i} data in the LMC taken with The Australia Telescope Compact Array (ATCA) / Parkes and discovered signatures of collision of H{\sc i} gas having two velocity components. The two components, the L and D components (\cite{1991IAUS..148...63R}, \yearcite{1993LNP...416..115R}; \cite{1992A&A...263...41L}), are separated by 60 km s$^{-1}$ in projection, and the collision signatures are the complementary distribution between the two components and the bridge features connecting them in velocity. It is suggested that the cluster RMC136 (hereafter, R136), which is most massive in the Local Group, and about 400 OB stars in the H{\sc i} ridge of the LMC were formed by the kpc-scale trigger in the last 10 Myr. N-body simulations by \citet{2007MNRAS.381L..16B} reproduced the H{\sc i} ridge and offer a theoretical basis supporting the trigger driven by the interaction. Subsequently, \citet{2019ApJ...871...44T} discovered that the high-mass stars in LHA 120-N 44 (hereafter, N44) was also formed by similar colliding H{\sc i} flows, and \citet{2019ApJ...886...14F} and \citet{2019ApJ...886...15T} presented additional pieces of evidence for high-mass star formation in the CO filamentary systems with Atacama Large Millimeter/submillimeter Array (ALMA), which are part of the same event. Because the galactic interaction in the Magellanic system is a unique event very close to the sun, the interaction is observed into unprecedented detail. It is suggested that the close encounter 200 Myr ago also drove the Magellanic Bridge connecting the SMC and the LMC \citep{2007MNRAS.381L..11M}. The SMC is known as an active galaxy in star formation (see \cite{2001PASJ...53L..45M} and references therein). It is important to explore if the tidal interaction triggers star formation in the SMC in an effort to better understand the origin of star formation in the Magellanic system. In the present work, we focus on NGC~602, an outstanding H{\sc ii} region with a young stellar cluster including 2000 $M_{\odot}$ in the southeast wing of the SMC.

The pre-main sequence stars in the center of NGC~602 was formed $\lesssim$5 Myr ago, and the recent star formation was triggered by stellar radiation $\sim$0.7--1.8 Myr ago at the edge of the H{\sc ii} region according to the previous works (\cite{2007ApJ...665L.109C}, \yearcite{2011ApJ...730...78C}; see also \cite{2009AJ....137.3668C,2012ApJ...748...64G}). In addition to NGC~602, there are more than ten O stars cataloged in the region within 500 pc of NGC~602.

The initial H{\sc i} results of the Galactic Australian Square Kilometre Array Pathfinder (GASKAP) have been published by \citet{2018NatAs...2..901M} and demonstrated the high power of this new facility. The present work is one of the efforts to pursuit star formation in the SMC by analyzing the high-resolution H{\sc i} data taken by the GASKAP. The present paper is organized as follows. Section \ref{sec:obs} give a description on the H{\sc i} datasets, and Section \ref{sec:results} explains the results of the analysis. In Section \ref{sec:discuss} we discuss the implications of the results of the H{\sc i} gas disruption and its relationship to the cluster formation. In Section \ref{sec:conc} conclusions are given.

\section{Observations}\label{sec:obs}
\subsection{ASKAP}
The GASKAP project is carried out as a test run of the Australian Square Kilometre Array Pathfinder (ASKAP) and covers the galactic latitude $\pm$$10{^\circ}$ of the Milky Way and the Large and Small Magellanic Clouds. The present data are part of the SMC data of the 21 cm H{\sc i} emission of the SMC \citep{2018NatAs...2..901M}. The angular resolution is 35\arcsec~$\times$ 27\arcsec~which corresponds to $\sim$10 pc at the distance of the SMC 60 kpc. The rms noise fluctuations are about 0.91 K at a velocity resolution of 3.9 km s$^{-1}$. 

\subsection{ALMA}
To derive physical properties of molecular clouds associated with NGC~602, we used archival ALMA Band 6 (211--275 GHz) CO datasets obtained in the Cycle 4 semester (Proposal \#2016.1.00360.S, PI: R. Indebetouw). The observations used the mosaic-mapping mode, covering a rectangle region of 198\arcsec~$\times$ 144\arcsec~centered at $\alpha_\mathrm{J2000}$ = $1^{\mathrm{h}}29^{\mathrm{m}}27\fs326$, $\delta_\mathrm{J2000}$ = $-73{^\circ}34\arcmin10\farcs523$. In the present study, we used the data obtained with the Atacama Compact Array (ACA): 8--10 antennas of the 7 m array as interferometer and 4 antennas of the total power (TP) array. The baseline ranged from 8.9 to 48.0 m, corresponding to $\it{u}$--$\it{v}$ distances from 6.8 to 36.8 k$\lambda$. The observation was conducted on October 2016 through four spectral windows. The target molecular lines were $^{12}$CO($J$ = 2--1), $^{13}$CO($J$ = 2--1), and C$^{18}$O($J$ = 2--1) with a bandwidth of 250 MHz for each (122.1 kHz $\times$ 2048 channels). In the present study, we analyzed only $^{12}$CO($J$ = 2--1) line emission data. Two quasars (J0522--3627 and J0006--0623) and Uranus were observed as complex gains and flux calibrators, respectively. A quasar J0450--8101 was also observed as a phase calibrator.

The data reduction was performed using the Common Astronomy Software Applications package (CASA, version 5.6.0; \cite{2007ASPC..376..127M}). We applied the multi-scale CLEAN algorithm implemented in CASA using the natural weighting \citep{2008ISTSP...2..793C}. Finally, we combined the cleaned 7 m array data with the total power data using feather procedure implemented in CASA. The synthesized beam size is 5\farcs6 $\times$ 7\farcs6 with a position angle of $-$62.2 degrees, corresponding to a spatial resolution of $\sim$2 pc at the distance of the SMC. The typical noise fluctuation is $\sim$0.04 K for a velocity resolution of 0.2 km s$^{-1}$.

\section{Results}\label{sec:results}
\begin{figure}[ht]
\begin{center}
\includegraphics[width=100mm]{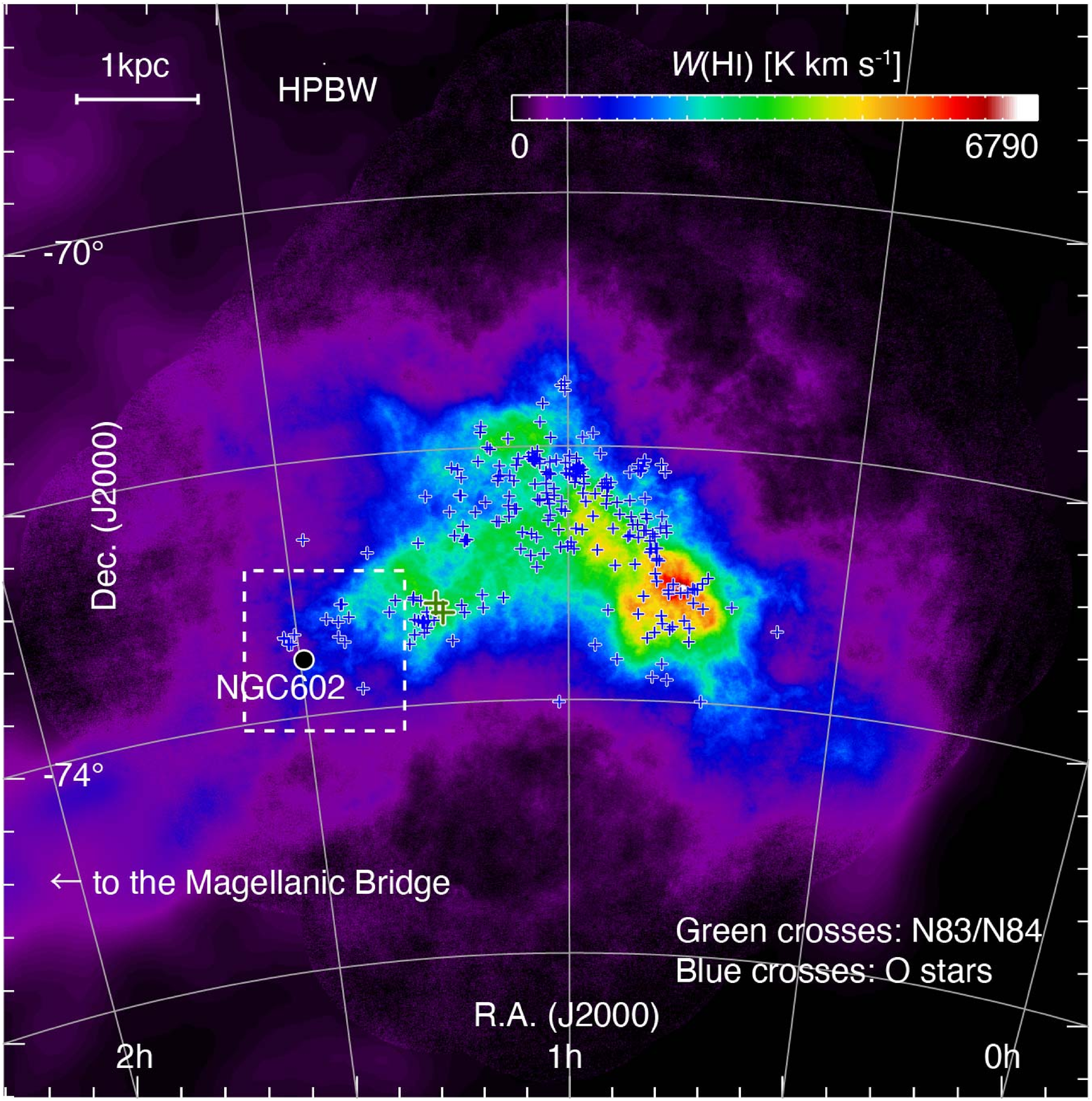}
\end{center}
\vspace*{0.5cm}
\caption{Integrated intensity map of H{\sc i} obtained with ASKAP \citep{2009IEEEP..97.1507D,2018NatAs...2..901M}. The integration velocity range is from 50.4 to 253.6 km s$^{-1}$. The filled circle indicates the position of NGC 602. The blue and green crosses represent the positions of O stars and H{\sc ii} regions N83 / N84, respectively. The beam size and scale bar are also indicated in the top left corner. The white dashed box represents the focused region in the present paper.}
\label{fig1}
\end{figure}

\begin{figure}[ht]
\begin{center}
\includegraphics[width=100mm]{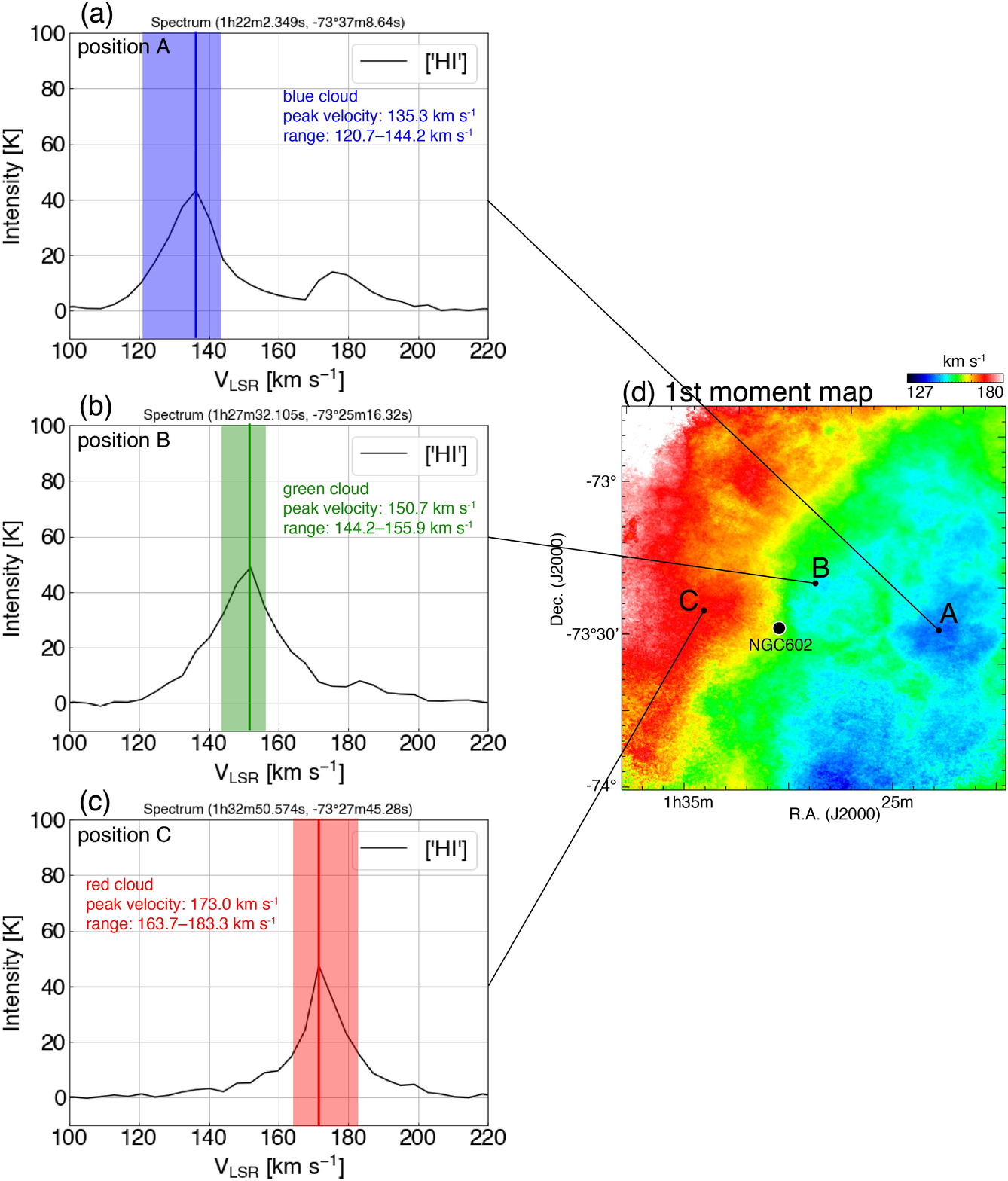}
\end{center}
\vspace*{0.5cm}
\caption{Typical H{\sc i} line profiles at position (a) A [$\alpha_\mathrm{J2000}$ = $1^{\mathrm{h}}22^{\mathrm{m}}2\fs3$, $\delta_\mathrm{J2000}$ = $-73{^\circ}37\arcmin8\farcs6$], (b) B [$\alpha_\mathrm{J2000}$ = $1^{\mathrm{h}}27^{\mathrm{m}}32\fs1$, $\delta_\mathrm{J2000}$ = $-73{^\circ}25\arcmin16\farcs3$], and (c) C [$\alpha_\mathrm{J2000}$ = $1^{\mathrm{h}}32^{\mathrm{m}}50\fs6$, $\delta_\mathrm{J2000}$ = $-73{^\circ}27\arcmin45\farcs3$]. Their peak velocities are 136 km s$^{-1}$, 150 km s$^{-1}$, and 170 km s$^{-1}$, respectively. And their velocity ranges are 120.7--144.2 km s$^{-1}$, 144.2--155.9 km s$^{-1}$, and 163.7--183.3 km s$^{-1}$, respectively. (d) First moment map around NGC 602. The filled circle indicates the position of NGC 602.}
\label{fig2}
\end{figure}

\begin{figure}[ht]
\begin{center}
\includegraphics[width=180mm]{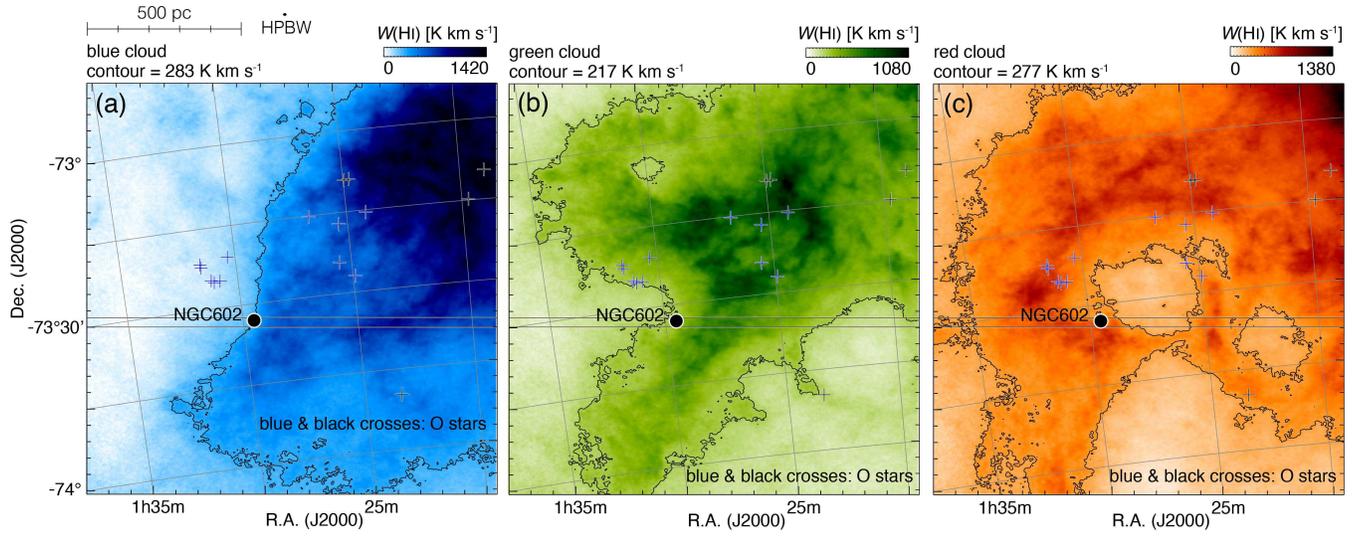}
\end{center}
\vspace*{0.5cm}
\caption{Integrated intensity maps of H{\sc i}. The velocity range is (a) 120.7--144.2 km s$^{-1}$ for the blue cloud, (b) 144.2--155.9 km s$^{-1}$ for the green cloud, and (c) 163.7--183.3 km s$^{-1}$ for the red cloud. The contour level is 20\% of the maximum integrated intensity for each panel. The filled circle is the same as those in Figure \ref{fig1}. The blue crosses represent positions of O stars discussed in the text and the black crosses represent positions of the other O stars. The black horizontal solid lines indicate the integration range of the right ascension-velocity diagram Figure \ref{fig4}. The beam size and scale bar are also indicated in the top left corner.}
\label{fig3}
\end{figure}

\begin{figure}[ht]
\begin{center}
\includegraphics[width=100mm]{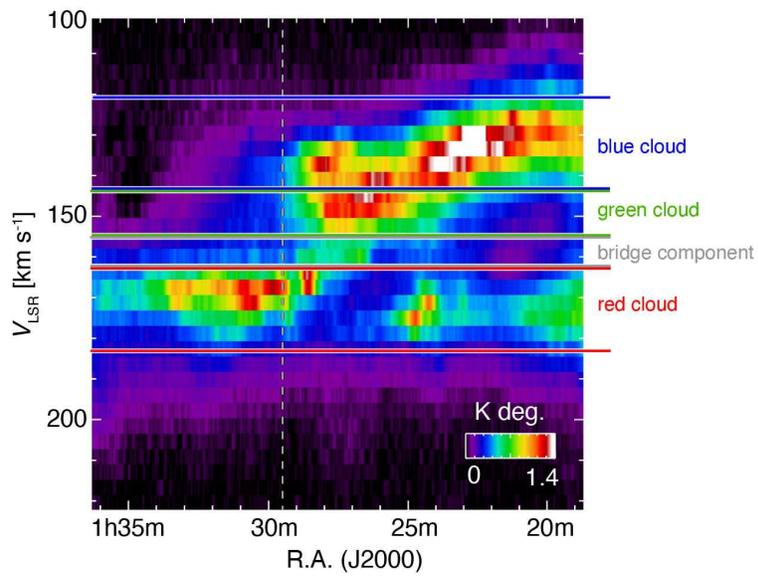}
\end{center}
\vspace*{0.5cm}
\caption{Right Ascension-velocity diagram of H{\sc i} around NGC602. The integration range of Declination is from $-73{^\circ}30\arcmin0\farcs0$ to $-73{^\circ}28\arcmin12\farcs0$. The blue, green, and red horizontal lines indicate the velocity ranges of the blue, green, and red clouds, respectively. The white dashed line indicates the position of NGC~602.}
\label{fig4}
\end{figure}

Figure \ref{fig1} shows the distribution of the total H{\sc i} integrated intensity of the 21 cm H{\sc i} emission $\it{W}$(H{\sc i}) [K km s$^{-1}$]. The open cluster NGC~602 is located toward the Southeastern wing of the SMC, while most of the O stars are distributed in the main body of the SMC. The analysis is made in the white box including NGC~602 in Figure \ref{fig1}. Figures \ref{fig2}(a), \ref{fig2}(b), and \ref{fig2}(c) show H{\sc i} line profiles at the three positions of the region of NGC~602, and Figure \ref{fig2}(d) shows the 1st moment distribution. We find three velocity components in Figure \ref{fig2}, and show the distribution of three components at a velocity range, defined by visual inspection using the right ascension-velocity diagram (Figure \ref{fig4}) and H{\sc i} profiles, of 120.7--144.2 km s$^{-1}$, 144.2--155.9 km s$^{-1}$ and 163.7--183.3 km s$^{-1}$, in Figures \ref{fig3}(a), \ref{fig3}(b), and \ref{fig3}(c), respectively. The blue-shifted gas (hereafter, the blue cloud) is in the southwest and the red-shifted gas (hereafter, the red cloud) is in the Northeast. The green-shifted gas (hereafter, the green cloud), in the intermediate velocity range, lies between the red and blue clouds. 

We pay attention to the peak of the green cloud which has a ring-like shape with a size of $\sim$25\arcmin. The red cloud has an intensity depression with a similar size to the green cloud and accompanies a ring-like shape of 30\arcmin--40\arcmin~diameter. The both ring-like features are somewhat elongated in the east-west direction. We also note that NGC~602 and the other eleven O stars \citep{2010AJ....140..416B} are distributed along the ring-like shape.

Table \ref{t1} gives the physical properties of the three clouds, the velocity range, the peak velocity, the peak H{\sc i} column density calculated in the optically thin approximation, the H{\sc i} mass, and the cloud radius defined as an effective radius $R$ calculated using the following equation;
\begin{equation}
R = (A / \pi)^{0.5},
\end{equation}
where $A$ is the area enclosed by a contour of 20\% of the maximum integrated intensity. We also used the expression for the H{\sc i} column density as follows; 
\begin{equation}
N(\rm {H\textsc{i}}) = 1.8224 \times 10^{18} \int_{\it v_{\rm {1}}}^{\it v_{\rm {2}}} \it {\Delta T_{\rm {b}} dv}~[\rm {cm^{-2}}]
\end{equation}
The H{\sc i} masses are (5--9)$\times$10$^{6}$ $M_{\odot}$ with a cloud radius of 500--600 pc. The H{\sc i} mass gives the lower limit and can be larger by a factor of $\sim$1.3 if the optical depth correction is applied (\cite{2014ApJ...796...59F}, \yearcite{2015ApJ...798....6F}, \yearcite{2018ApJ...860...33F}). A fuller account of the H{\sc i} optical depth in the SMC is a subject of a future work.

Figure \ref{fig4} shows a position-velocity diagram taken along R.A. The diagram shows the three velocity components as indicated by the two colored straight lines at constant Dec. In the diagram we see connecting features in velocity between the green and red clouds at $\alpha_\mathrm{J2000}$ = $1^{\mathrm{h}}26^{\mathrm{m}}$--$1^{\mathrm{h}}29^{\mathrm{m}}$ in a velocity range of 155.9--163.7 km s$^{-1}$. It seems that the blue and green clouds are continuous in velocity, suggesting that they are a single body with a velocity gradient due to the rotation of the SMC \citep{2004ApJ...604..176S,2019MNRAS.483..392D}, while the red cloud is clearly separated from the others. Because we focus on the localized area at the edge of the SMC, we applied no correction for the rotation. This does not cause a problem in the present study.

\begin{figure}[ht]
\begin{center}
\includegraphics[width=150mm]{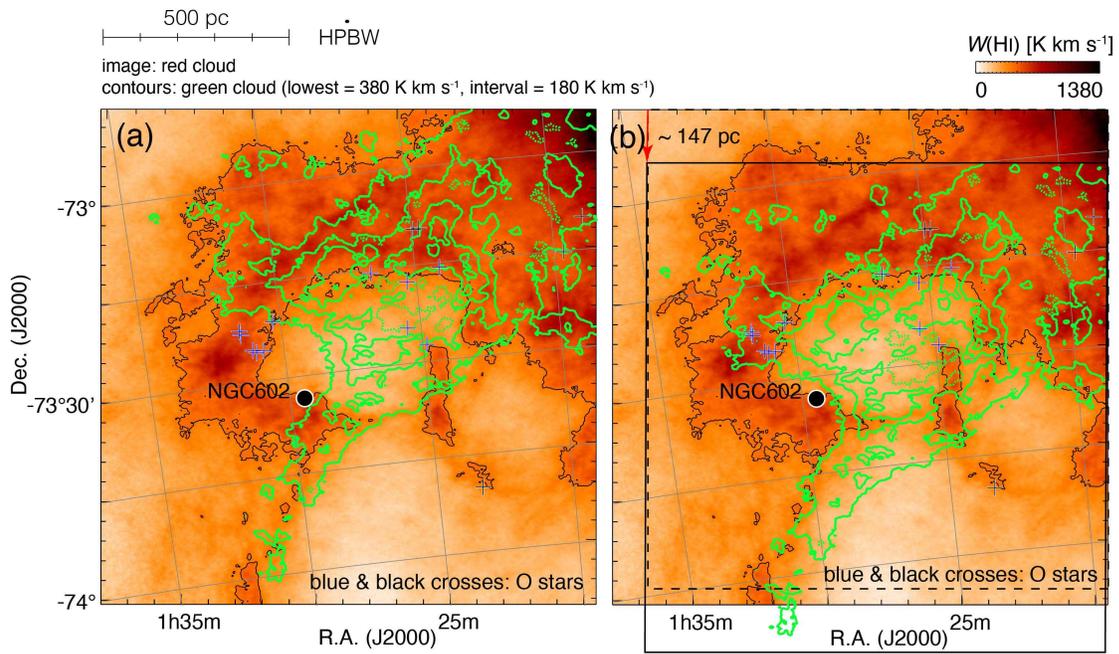}
\end{center}
\vspace*{0.5cm}
\caption{(a) Comparison between gas distributions of the green cloud (green contours) and the red cloud (image and black contour). The velocity ranges of the green and red clouds are the same as shown in Figures \ref{fig3}(b) and \ref{fig3}(c), respectively. The contour levels of green cloud are 380, 560, 740, and 920 K km s$^{-1}$. (b) Complementary spatial distribution of the green cloud (contours) and the red cloud (image). The contours are spatially displaced $\sim$147 pc in the direction of south. The dashed and solid boxes indicate before and after displacement of the contours, respectively. The filled circle is the same as those in Figure \ref{fig1}. The blue and black crosses are the same as those in Figure \ref{fig3}. The beam size and scale bar are also indicated in the top left corner.}
\label{fig5}
\end{figure}

Figure \ref{fig5}(a) shows an overlay of the green and red clouds. It shows that the green cloud and the ring-like intensity depression of the red cloud have complementary distribution with each other. We find some displacement between the two clouds and applied the algorithm developed to trace a collision signature by \citet{2020arXiv200313925F} (see also for supplementary explanation \cite{2018ApJ...859..166F}). Figure \ref{fig5}(b) shows an overlay with the displacement vector of 147 pc in the north to the south direction, and we find the two ring-like features corresponds well with each other. Furthermore, the H{\sc i} tails extending toward the southeast in the two clouds also show complementary distribution.

\begin{figure}[ht]
\begin{center}
\includegraphics[width=100mm]{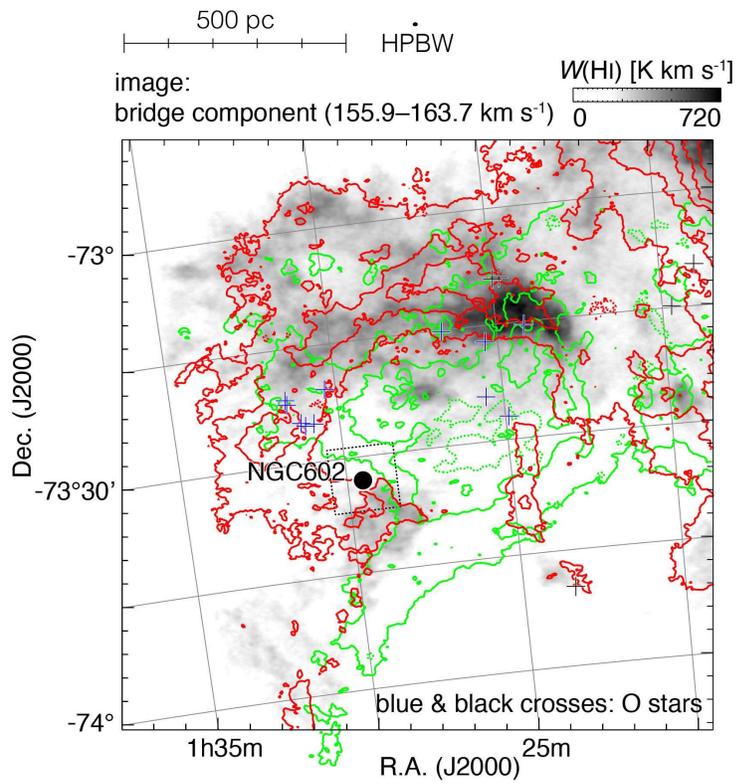}
\end{center}
\vspace*{0.5cm}
\caption{Comparison among gas distributions of the green cloud (green contours), the red cloud (red contours), and the bridge component (image). The velocity ranges of the green and red clouds are the same as shown in Figures \ref{fig3}(b) and \ref{fig3}(c), and image (the bridge component) is 155.9--163.7 km s$^{-1}$, respectively. The filled circle is the same as those in Figure \ref{fig1}. The blue and black crosses are the same as those in Figure \ref{fig3}. Focused region in Figure \ref{fig7}(a) is indicated by the black dashed box. The beam size and scale bar are also indicated in the top left corner.}
\label{fig6}
\end{figure}

Figure \ref{fig6} shows another overlay of the two clouds with the distribution of the bridge component in the position-velocity diagram (Figure \ref{fig4}). We see corresponding distribution of the bridge feature toward the ring-like distribution of the red cloud.

\begin{figure}[ht]
\begin{center}
\includegraphics[width=100mm]{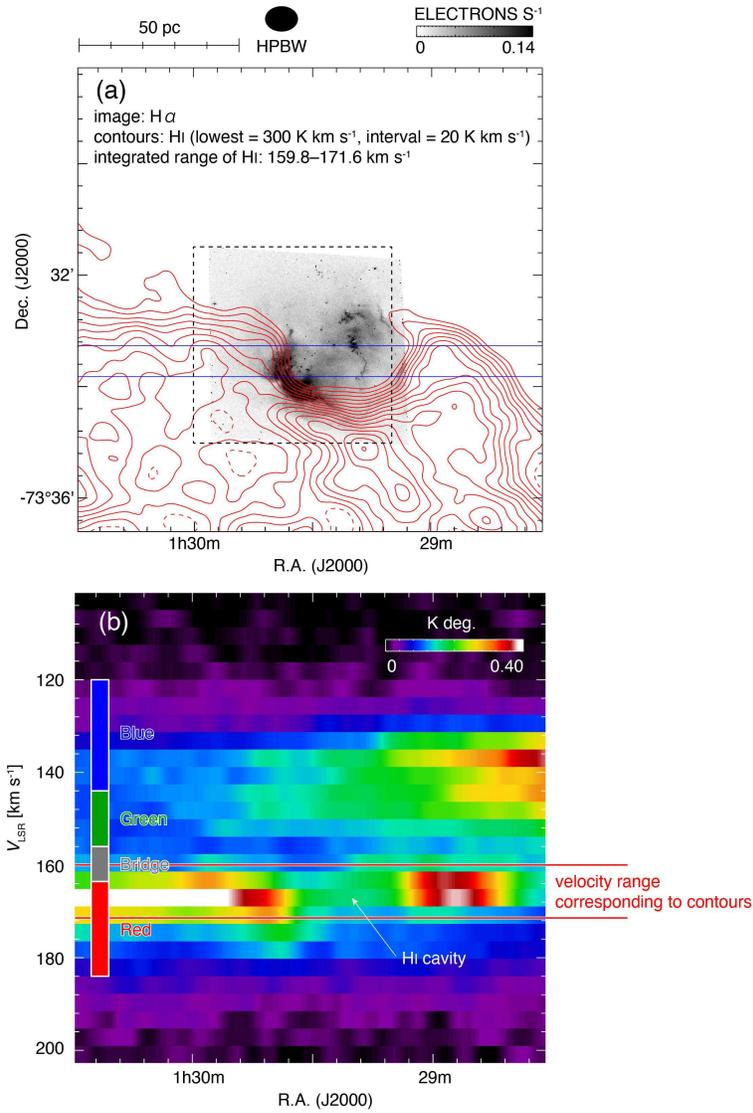}
\end{center}
\vspace*{0.5cm}
\caption{(a) H{\sc i} integrated intensity contours superposed on the optical image of NGC~602. The velocity range of  H{\sc i} is 159.8--171.6 km s$^{-1}$. The lowest contour level and contour interval are 300 K km s$^{-1}$ and 20 K km s$^{-1}$, respectively. The blue horizontal solid lines indicate the integration range of the right ascension-velocity diagram Figure 7(b). Focused region in Figure \ref{fig8} is indicated by the black dashed box. The beam size and scale bar are also indicated in the top left corner. (b) Right Ascension-velocity diagram of H{\sc i}. The integration range of Declination is from $-73{^\circ}33\arcmin36\farcs0$ to $-73{^\circ}33\arcmin0\farcs0$. The red horizontal solid lines indicate the H{\sc i} integration velocity range in Figure 7(a). The blue, green, gray, and red strips indicate the velocity ranges of the blue cloud, the green cloud, the bridge component, and red clouds, respectively.}
\label{fig7}
\end{figure}

Figure \ref{fig7}(a) shows a close up view the H{\sc i} toward NGC~602 which shows an intensity depression of H{\sc i} toward the H{\sc ii} region. The coincidence was previously suggested at lower resolution by \citet{2008PASP..120..972N}. In particular, the southern half of NGC~602 shows a good correspondence with the H{\sc i} cavity of 50 pc diameter over a velocity range of 159.8--171.6 km s$^{-1}$ at the present high spatial resolution, which includes part of the red cloud and the bridge feature.

Figure \ref{fig8} shows a contour map of $^{12}$CO($J$ = 2--1) obtained using ALMA superposed on the $HST$ optical composite image toward NGC~602. CO distribution is highly clumpy. We identified 19 clouds whose mass and cloud radius ranges are 10--2000 $M_{\odot}$ and 10--50 pc. Criteria of cloud identification and their physical properties are shown in Appendix. All the clouds are tightly connected with the boundary of the H{\sc ii} region which shows bright rim. The mass of the CO clouds is shown in Table \ref{t2}, and the total mass of the CO clouds are estimated to be $\sim$3800 $M_{\odot}$.

\section{Discussion} \label{sec:discuss}
Following the collision scenario proposed for the LMC \citep{2017PASJ...69L...5F,2019ApJ...871...44T,2019ApJ...886...14F,2019ApJ...886...15T}, we present a scenario that the two H{\sc i} clouds collided with each other to trigger the formation of NGC~602 and the other eleven O stars in the region. The displacement 147 pc between the two clouds in Figure \ref{fig7} shows that the red cloud possibly moved from the north and collided with the green cloud to create the cavity in the red cloud. It is probable that the collision triggered the formation of the high-mass stars including NGC~602 and the eleven O stars in the red cloud.

The velocity range of the red cloud is 159.8--171.6 km s$^{-1}$ and that of the CO clouds is 158.8--178.8 km s$^{-1}$. Thus, the velocity range is common both in the H{\sc i} (the red cloud) and CO. We infer that the ring-like distributions in the red cloud and the green cloud were formed by gas compression driven by the collision. A possible scenario may be that the collision produced shocks in them which are propagating at $\sim$10 km s$^{-1}$ both outward in the red cloud and inward in the green cloud, while detailed hydrodynamical process remains to be carried out by numerical simulations of hydrodynamics.

The time scale of the collision is roughly estimated to be 7.7 Myr from a ratio of the displacement 147 pc and the relative velocity 20 km s$^{-1}$ by assuming an angle of the collision vector to the line of sight to be 45 degrees. The age of NGC~602 is estimated from optical studies to be less than $\sim$5 Myr. This is consistent with the collision duration in the present scenario. We assume that the collision took place between the H{\sc i} gas with no CO. It is likely that the compressed H{\sc i} gas formed H$_{2}$ in a time scale of Myr as shown by the MHD simulations of colliding H{\sc i} flows (\cite{2012ApJ...759...35I,2018ApJ...860...33F,2018arXiv181102224T}; R. Maeda et al. in preparation). Star formation may be continuing at present in the CO clouds, whose mass is large enough to form more high-mass stars in future. We suggest that the collision compressed H{\sc i} gas with H{\sc i} column density of 1.5 $\times$ 10$^{21}$ cm$^{-2}$. The H{\sc i} mass in the cavity is roughly estimated to be $\sim$10$^{4}$ $M_{\odot}$ assuming the uniform density of 1.5 $\times$ 10$^{21}$ cm$^{-2}$ within the H{\sc i} cavity, which is large enough to form the CO clouds of $\sim$3800 $M_{\odot}$ and the stars of 2000 $M_{\odot}$ by compression. The star formation efficiency may be estimated to be $\sim$20\% from a ratio of the stellar mass and the H{\sc i} mass, 2000 $M_{\odot}$ / 10$^{4}$ $M_{\odot}$. Therefore, CO clouds may be detected toward the eleven O stars.

The origin of the Magellanic Bridge is not yet fully understood. \citet{2007MNRAS.381L..11M} presented numerical simulations of the close encounter between the LMC and the SMC and showed that the Magellanic Bridge can be formed as a result of the interaction. The observed H{\sc i} and CO velocity of the Magellanic Bridge ranges from 140 to 180 km s$^{-1}$, which is similar to that of the red cloud (\cite{2003MNRAS.338..609M}, \yearcite{2003MNRAS.339..105M}, \yearcite{2004ApJ...616..845M}; \cite{2006ApJ...643L.107M}). 

It has been known that the SMC has many shell like features in H{\sc i} \citep{1980MNRAS.192..365M,1997yCat..72890225S,1999MNRAS.302..417S,2005MNRAS.360.1171H}, and it was suggested that multiple H{\sc i} shells colliding in the region of NGC~602 triggered the formation of the cluster \citep{2008PASP..120..972N}. Usually, it is thought that H{\sc i} shells are driven by stars, whereas in many shells the driving stars are not identified. The present study shows that the H{\sc i} shell in the red cloud was created by the collision with another H{\sc i} cloud. This is a similar case to the H{\sc i} supergiant shell LMC~2, which turned out to have been formed by a collision between H{\sc i} flows at kpc scale which triggered the formation of R136 and N159E and W (\cite{2015ApJ...807L...4F}, \yearcite{2017PASJ...69L...5F}, \yearcite{2019ApJ...886...14F}; \cite{2017ApJ...835..108S,2019ApJ...886...15T}).

\section{Conclusions} \label{sec:conc}
We have analyzed the new GASKAP H{\sc i} data toward NGC~602 in the SMC and discovered observational signatures for cloud-cloud collision which triggered the formation of NGC~602 and the other eleven O stars. The main conclusions are summarized as follows;

\begin{enumerate}
\item \label{enu1} The H{\sc i} gas in the NGC~602 region is in a velocity range of 110--200 km s$^{-1}$ and has three velocity components peaked at 135.3 km s$^{-1}$, 150.7 km s$^{-1}$, and 173.0 km s$^{-1}$. The clouds at three velocities show significantly different distribution with each other, in particular, the red cloud shows different distributions from the other two. We find that the red cloud has a cavity, which was previously identified as a supershell \citep{1997yCat..72890225S,1999MNRAS.302..417S}, showing complementary distribution with the green cloud. The green cloud and the cavity in the red cloud matches well with each other by displacing the green cloud to the south by 147 pc. Such a displacement is a typical signature of colliding clouds. In addition, the two clouds are linked by the bridge features in velocity. 
\item \label{enu2} Based on the signatures in \ref{enu1}, we present a scenario that the two clouds collided with each other in the past. NGC~602 and the O stars are distributed along the cavity in the red cloud, and we interpret that the collision triggered the formation of NGC~602 and the O stars and the two ring-like H{\sc i} distributions of 150 pc radius in the green cloud and 200 pc radius in the red cloud. The typical time scale of the collision is estimated to be $\sim$8 Myr from a ratio of the displacement 150 pc and the velocity 20 km s$^{-1}$, while the duration of the collisional compression rages over a longer time scale due to the cloud size. The age of NGC~602 1--5 Myr is consistent with the timescale if the collision continued over 8 Myr by compressing the H{\sc i} gas to form CO clouds and the stars. The location of NGC~602 is in the southern edge of the cavity and it is likely that the compression happened in the earliest phase of the collision. The collision is continuing now in the northern part of the cavity where dense H{\sc i} gas is located. The green cloud is moving from the south, i.e., the location of the Magellanic Bridge. 
\item \label{enu3} NGC~602 is associated with the red-shifted velocity component and we find a clear H{\sc i} cavity of 50 pc diameter associated with the H{\sc ii} region. A recent ALMA CO image indicates near twenty clumpy CO clouds whose total mass is $\sim$3800 $M_{\odot}$, which is comparable to the stellar mass 2000 $M_{\odot}$. The total H{\sc i} mass of the red cloud within the H{\sc ii} region is estimated to be $\sim$10$^{4}$ $M_{\odot}$ and the star formation efficiency becomes 20\% if the H{\sc i} cloud is assumed as the parent cloud.
\end{enumerate}

\begin{ack}
The authors are grateful to Kenji Bekki for his insightful comments on the tidal interaction in the Magellanic system, which helped to improve the manuscript. The Australian SKA Pathfinder is part of the Australia Telescope National Facility which is managed by CSIRO. Operation of ASKAP is funded by the Australian Government with support from the National Collaborative Research Infrastructure Strategy. ASKAP uses the resources of the Pawsey Supercomputing Centre. Establishment of ASKAP, the Murchison Radio-astronomy Observatory and the Pawsey Super- computing Centre are initiatives of the Australian Government, with support from the Government of Western Australia and the Science and Industry Endowment Fund. We acknowledge the Wajarri Yamatji people as the traditional owners of the Observatory site. This paper makes use of the following ALMA data: ADS/JAO.ALMA\#2016.1.00360.S. ALMA is a partnership of ESO (representing its member states), NSF (USA) and NINS (Japan), together with NRC (Canada), MOST and ASIAA (Taiwan), and KASI (Republic of Korea), in cooperation with the Republic of Chile. The Joint ALMA Observatory is operated by ESO, AUI/NRAO and NAOJ. In addition, publications from NA authors must include the standard NRAO. acknowledgement: The National Radio Astronomy Observatory is a facility of the National Science Foundation operated under cooperative agreement by Associated Universities, Inc. Based on observations made with the NASA/ESA Hubble Space Telescope, and obtained from the Hubble Legacy Archive, which is a collaboration between the Space Telescope Science Institute (STScI/NASA), the Space Telescope European Coordinating Facility (ST-ECF/ESA) and the Canadian Astronomy Data Centre (CADC/NRC/CSA). This work was financially supported in part by JSPS KAKENHI Grant Numbers 15H05694, 19K14758, and 19H05075.
$Software:$ CASA (v 4.5.3.: \cite{2007ASPC..376..127M})
\end{ack}
\clearpage

\begin{figure}[ht]
\begin{center}
\includegraphics[width=100mm]{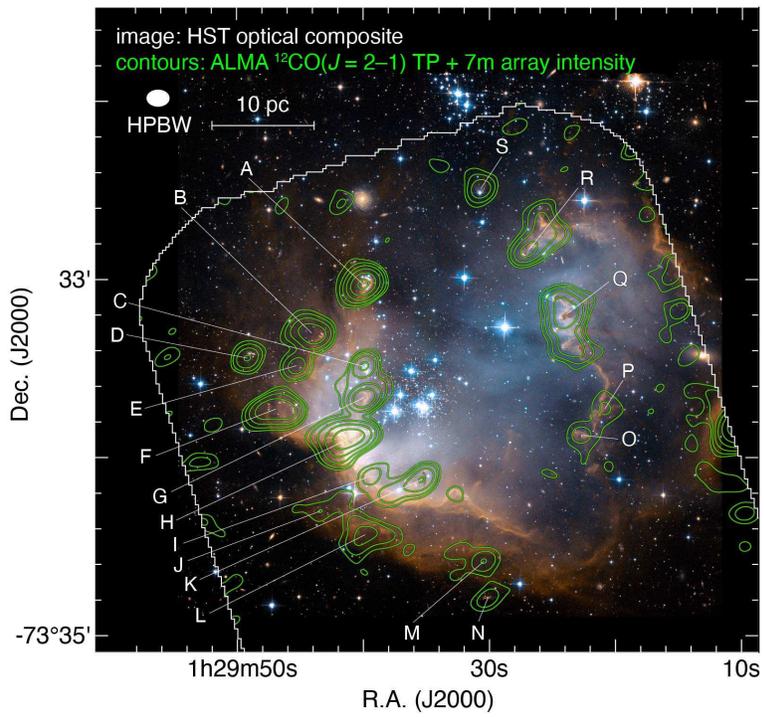}
\end{center}
\vspace*{0.5cm}
\caption{Overlay map of $\it{HST}$ optical image [credit: NASA, ESA, and the Hubble Heritage Team (STScI/AURA)-ESA/Hubble Collaboration] and $^{12}$CO($J$ = 2--1) integrated intensity contours obtained with ALMA. The contour levels are 0.4 (5$\sigma$), 0.8, 1.6, 3.2, 6.4, and 12.8 K km s$^{-1}$. The integration range is 158.8--178.8 km s$^{-1}$. The CO peaks A--S discussed in {\S}\ref{sec:discuss} are indicated. An observed area is indicated with white solid line, and the beam size and scale are also indicated in the top left corner.}
\label{fig8}
\end{figure}

\begin{appendix}
\begin{center}
APPENDIX\\
Physical properties of molecular clouds
\end{center}
To derive the physical properties of the molecular clouds associated with NGC~602, we identified CO clouds according to the following criteria:
\begin{enumerate}
\item An area enclosed by the contour of 0.4 K km s$^{-1}$ (equal to 5$\sigma$ level for the integration for a velocity range of $V$$_{\rm {LSR}}$ =158.8--178.8 km s$^{-1}$) in Figure \ref{fig8} is defined as an individual cloud.
\item When the major axis of an area is smaller than 10\arcsec~(the beam size is $7\farcs3$), the area is not defined as an individual cloud.  
\item When there are multiplex peaks in the same area, the peaks are divided by the minimum in the intensity map and are defined as independent.
\end{enumerate}
As a result, we identified 19 molecular clouds as shown in Figure \ref{fig8}.

The CO-derived mass $M$$_{\rm {CO}}$ is obtained from the following equations:
\begin{equation}
M_{\rm {CO}} = m_{\rm H} \mu \sum_{i} [D^2 \Omega N_i(\rm {H_2})]
\end{equation}
and
\begin{equation}
N_{\rm {H_2}} = X_{\rm {CO}} W(\rm {CO}),
\end{equation}
where $m_{\rm H}$ is the mass of atomic hydrogen, $\mu$ is the mean molecular weight relative to atomic hydrogen, $D$ is the distance to the source in units of cm, $\Omega$ is the solid angle subtended by a unit grid, $N_i(\rm {H_2})$ is the molecular hydrogen column density for each pixel in units of cm$^{-2}$, $X_{\rm {CO}}$ is CO-to-H$_{2}$ conversion factor, and $W$(CO) is the $^{12}$CO($J$ = 1--0) integrated intensity. We applied equation (A1) for the area enclosed by a contour of 50\% of the maximum integrated intensity. We adopted $\mu = 2.7$ and $X_{\rm {CO}} = 7.5 \times 10^{20}$ cm$^{-2}$ (K km s$^{-1}$)$^{-1}$ \citep{2017ApJ...844...98M}. We calculated $W$(CO) from our $^{12}$CO($J$ = 2--1) data using $^{12}$CO($J$ = 2--1) / ($J$ = 1--0) ratio of 1.0 according to \citet{1996A&AS..118..263R}. This ratio of 1.0 is derived from the average value of whole the previous observations of molecular clouds in the SMC (LIRS~49, LIRS~36, SMC-B1\#1, N66, N88, and Hodge15).
\end{appendix}
\clearpage

\begin{table}[ht]
\tbl{Physical properties of H{\sc i} clouds}{%
\begin{tabular}{lccccc} \hline \hline
Cloud name & Velocity range & Peak velocity & Peak column density & Mass & Cloud radius \\
& [km s$^{-1}$] & [km s$^{-1}$] & [$\times$10$^{21}$ cm$^{-2}$] & [$\times$10$^{6}$ $M_{\odot}$] & [pc] \\
(1) & (2) & (3) & (4) & (5) & (6) \\ \hline
Blue cloud & 120.7--144.2 & 135.3 & 2.6 & 9.1 & 600\\
Green cloud & 144.2--155.9 & 150.7 & 1.7 & 4.9 & 500\\
Bridge component & 155.9--163.7 & ----- & 1.3 & 4.0 & 500\\
Red cloud & 163.7--183.3 & 173.0 & 2.5 & 7.6 & 600\\ \hline
\end{tabular}}
\begin{tabnote}
\hangindent6pt\noindent
Note. --- Col. (1): Cloud name. Col. (2) Velocity range of each cloud defined by visual inspection. Col. (3) Peak velocity of each cloud derived by Gaussian fitting. Col. (4): Peak column density of eacd cloud calculated using equation (2) Col. (5): Mass of each cloud calculated using equation (2). Col. (6): Cloud radius defined as an effective radius $R$ calculated using equation (1).
\end{tabnote}
\label{t1}
\end{table}
\clearpage

\begin{table}[ht]
\tbl{Physical properties of CO clouds}{%
\begin{tabular}{lccccccc} \hline \hline
Object & R.A. & Dec. & \it{T}$_{\rm peak}$ & \it{V}$_{\rm peak}$ & $\Delta V$ & \it{R} & \it{M}$_{\rm CO}$ \\ 
& [hms] & [dms] & [K] & [km s$^{-1}$] & [km s$^{-1}$] & [pc] & [$M_{\odot}$] \\
(1) & (2) & (3) & (4) & (5) & (6) & (7) & (8) \\ \hline
A & $1^{\mathrm{h}}29^{\mathrm{m}}35\fs8$ & $-73{^\circ}32\arcmin60\farcs0$ & 4.3 & 173.0 & 3.3 & 2.1 & 580\\
B & $1^{\mathrm{h}}29^{\mathrm{m}}39\fs6$ & $-73{^\circ}33\arcmin18\farcs0$ & 3.1 & 171.2 & 1.7 & 2.6 & 320\\
C & $1^{\mathrm{h}}29^{\mathrm{m}}35\fs5$ & $-73{^\circ}33\arcmin28\farcs8$ & 2.3 & 173.8 & 1.4 & 2.0 & 130\\
D & $1^{\mathrm{h}}29^{\mathrm{m}}44\fs9$ & $-73{^\circ}33\arcmin25\farcs2$ & 2.1 & 166.3 & 1.6 & 1.9 & 120\\
E & $1^{\mathrm{h}}29^{\mathrm{m}}41\fs0$ & $-73{^\circ}33\arcmin32\farcs4$ & 2.2 & 171.1 & 1.0 & 2.3 & 120\\
F & $1^{\mathrm{h}}29^{\mathrm{m}}43\fs0$ & $-73{^\circ}33\arcmin43\farcs2$ & 4.2 & 166.5 & 1.3 & 3.0 & 530\\
G & $1^{\mathrm{h}}29^{\mathrm{m}}35\fs5$ & $-73{^\circ}33\arcmin39\farcs6$ & 1.9 / 0.7 & 175.0 / 177.4 & 2.5 / 1.7 & 2.0 & 240\\
H & $1^{\mathrm{h}}29^{\mathrm{m}}37\fs4$ & $-73{^\circ}33\arcmin54\farcs0$ & 3.7 & 169.3 & 2.0 & 2.7 & 590\\
I & $1^{\mathrm{h}}29^{\mathrm{m}}35\fs4$ & $-73{^\circ}34\arcmin05\farcs0$ & 0.5 & 170.9 & 2.1 & 2.5 & \phantom{0}70\\
J & $1^{\mathrm{h}}29^{\mathrm{m}}39\fs2$ & $-73{^\circ}34\arcmin18\farcs2$ & 0.7 & 167.0 & 1.1 & 2.1 & \phantom{0}40\\
K & $1^{\mathrm{h}}29^{\mathrm{m}}31\fs0$ & $-73{^\circ}34\arcmin04\farcs8$ & 1.5 & 164.7 & 1.7 & 2.1 & 100\\
L & $1^{\mathrm{h}}29^{\mathrm{m}}35\fs8$ & $-73{^\circ}34\arcmin26\farcs4$ & 0.8 / 0.5 & 166.5 / 164.6 & 1.6 & 2.1 & \phantom{0}40\\
M & $1^{\mathrm{h}}29^{\mathrm{m}}26\fs3$ & $-73{^\circ}34\arcmin34\farcs7$ & 0.9 & 166.0 & 1.8 & 2.0 & \phantom{0}60\\
N & $1^{\mathrm{h}}29^{\mathrm{m}}25\fs8$ & $-73{^\circ}34\arcmin46\farcs8$ & 1.6 & 166.3 & 0.8 & 2.0 & \phantom{0}50\\
O & $1^{\mathrm{h}}29^{\mathrm{m}}18\fs5$ & $-73{^\circ}33\arcmin54\farcs0$ & 0.4 & 163.0 & 2.5 & 2.5 & \phantom{0}60\\
P & $1^{\mathrm{h}}29^{\mathrm{m}}16\fs9$ & $-73{^\circ}33\arcmin44\farcs1$ & 0.4 & 161.7 & 2.1 & 2.8 & \phantom{0}60\\
Q & $1^{\mathrm{h}}29^{\mathrm{m}}19\fs4$ & $-73{^\circ}33\arcmin10\farcs8$ & 2.5 & 161.2 & 2.3 & 2.7 & 390\\
R & $1^{\mathrm{h}}29^{\mathrm{m}}22\fs8$ & $-73{^\circ}32\arcmin49\farcs2$ & 2.0 & 162.6 & 1.8 & 2.3 & 190\\
S & $1^{\mathrm{h}}29^{\mathrm{m}}26\fs6$ & $-73{^\circ}32\arcmin31\farcs2$ & 1.9 & 167.7 & 1.4 & 2.0 & 110\\ \hline
\noalign{\vskip3pt} 
\end{tabular}}
\begin{tabnote}
\hangindent6pt\noindent
Note. --- Col. (1): Cloud name. Cols. (2)--(3): Position of the maximum integrated intensity of $^{12}$CO($J$ = 2--1) for each cloud. Cols. (4)--(6): Physical properties of the $^{12}$CO($J$ = 2--1) spectra for each CO cloud. Col. (4): Peak radiation temperature $T_{\rm {peak}}$. Col. (5): Center velocity $V_{\rm {peak}}$ derived by Gaussian fitting. Col. (6): FWHM line width $\Delta V$. Col. (7): Cloud radius defined as an effective radius $R$ = (A/$\pi$)$^{0.5}$, where A is the area enclosed by a contour of 50\% of the maximum integrated intensity. Col. (8): Mass of each cloud calculated using $\it{N}$(H$_{\rm 2}$) = 7.5 $\times$ 10$^{20}$ $W(^{12}\rm {CO})$ cm$^{-2}$ (K km s$^{-1}$)$^{-1}$, where $\it{N}$(H$_{\rm 2})$ is the molecular hydrogen column density and $\it{W}$($^{12}\rm CO$) is the $^{12}$CO($J$ = 1--0) intensity \citep{2017ApJ...844...98M}. We assume the intensity ratio of $^{12}$CO ($J$ = 2--1) / ($J$ = 1--0) = 1.0 (see the text).
\end{tabnote}
\label{t2}
\end{table}
\clearpage

\clearpage

\clearpage

\clearpage

\clearpage

\clearpage

\clearpage

\clearpage

\end{document}